\documentstyle[prl,preprint,aps]{revtex}
\tightenlines
\begin{document}
 
\draft
\title{Moment scaling at the sol - gel transition}

\author
{Robert Botet$^{\dagger}$ and Marek P{\l}oszajczak$^{\ddagger}$}
 
\address{$^{\dagger}$
 Laboratoire de Physique des Solides - CNRS,
B\^{a}timent 510, Universit\'{e} Paris-Sud
\\ Centre d'Orsay, F-91405 Orsay, France  \\
and   \\  $^{\ddagger}$
Grand Acc\'{e}l\'{e}rateur National d'Ions Lourds (GANIL), \\
CEA/DSM -- CNRS/IN2P3, BP 5027,  F-14021 Caen Cedex, France  }
 
\date{\today}
 
\maketitle
 
\begin{abstract}
\parbox{14cm}{\rm 
Two standard models of sol-gel transition are revisited here
from the point of view of their fluctutations in various  moments of both the
mass-distribution  and the gel-mass. Bond-percolation model 
is an at-equilibrium system and undergoes a static second-order
phase transition, while Monte-Carlo Smoluchowski model
is an off-equilibrium one and shows a dynamical critical phenomenon.
We show that the macroscopic quantities can be splitted into the three
classes with different scaling properties of their fluctuations, depending on
wheather they correspond to : (i) non-critical quantities, (ii) critical
quantities or to (iii) an
order parameter. All these three scaling properties correspond to a single
form : $<M>^{\delta } P(M) = \Phi ((M-<M>)/<M>^{\delta } )$~, with the values of
$\delta $ respectively : $=1/2$ (regime (i)), $\neq 1/2$ and $1$ (regime (ii)),
 and $=1$ (regime (iii)).
These new scalings are very robust and, in particular, they do 
not depend on the precise form of an Hamiltonian.} 
 
\end{abstract}
\bigskip

\vfill
\newpage

\section{Reversible and irreversible aggregation models}

Below we deal with a system of monomers (the basic units) 
which can aggregate to form the connected clusters. A monomer is considered 
as a cluster
of mass 1 (this is the mass unit). Moreover, monodispersity of the monomeric
mass as well as the mass conservation during the 
aggregation is assumed, i.e., the cluster-masses are always integer numbers 
in this approach. 

The basic sol-gel transition is the appearance at a finite time of 
an infinite cluster, called {\it the gel}. 'Infinite' means here 
that a finite fraction of the total mass of the system belongs 
to the gel. Note that this definition is applicable both to
finite as well as to infinite systems, but contextually and historically,
the tools used to study this behaviour have been defined somewhat differently
in these two cases \cite{stauffer}. For finite systems, the following moments
of the number-mass-distribution $n_s$ ( i.e. {\it the number} of clusters
of mass $s$) are introduced :
\begin{eqnarray}
\label{eq1}
M_k' = \sum_{s} s^k n_s~~~ \ ,
\end{eqnarray}
where the summation is performed over all clusters except the largest one.
The superscript $'$~ in Eq. (\ref{eq1}) recalls this constraint on the allowed 
values of $s$~. Consequently, 
the mass of gel-fraction is just $N-M_1'$~ with :
\begin{eqnarray}
N = \sum_{{\text all~ s}} s n_s~~~ \ ,
\nonumber
\end{eqnarray} 
the total mass of the system.
For infinite systems, the following normalized moments of the
concentration-mass-distribution $c_s$ (i.e. {\it the concentration} of clusters
of mass $s$) are introduced : 
\begin{eqnarray}
\label{eq2}
m_k' = \sum_{s} s^k c_s~~~ \ ,
\end{eqnarray}
where the summation runs over all $s$~. 
Generally, the concentrations are normalized 
in a special way : 
\begin{eqnarray}
c_s= {\lim }_{N \rightarrow \infty} \frac{n_s}{N} ~~~ \ ,
\nonumber
\end{eqnarray} 
and not in a direct relation with the volume. Consequently,
the probability for a monomer to belong to the gel-cluster 
is just equal to $1-m_1'$~.

These different definitions of moments are equivalent in the sense that
they are connected through the relation : 
\begin{eqnarray}
m_k'= {\lim }_{N \rightarrow \infty} \frac{M_k'}{N} ~~~ \ .
\nonumber
\end{eqnarray} 

In this paper, we investigate scaling behaviours of 
distributions of mass-distribution-moments for the two 'classical'
approaches to the sol-gel transition : the reversible 
model (the percolation \cite{gennes}) and the irreversible model 
(the Smoluchowski equations\cite{smo}). In the former case, diffusion 
is unimportant
and the reaction between clusters is the relevant step, whereas in the latter
case,  on the contrary, the reaction between clusters is unimportant and
the diffusion is relevant. In this sense, 
these two models are believed to belong to the two different classes,
and all other models of aggregation are suspected to belong to one of
them. For example, statistical Flory-Stockmayer theory\cite{fs}
should behave as the percolation model for the scaling
properties, since it corresponds to the calculation 'at equilibrium'.
This '{\it universality}' in diverse sol-gel situations
is at present a guess.

Percolation model\cite{stauffer} can be defined as follows : in a box
(a part of the regular lattice), each site correponds to a monomer and 
a proportion $p$ of active bonds is set randomly between sites. 
This results in a distribution of clusters defined
as ensemble of sites connected by active bonds. For a definite 
value of $p$, say $p_{cr}$, a giant cluster almost surely 
spans all the box. For example, in the thermodynamic limit when the 
size of box becomes infinite (this limit will be
denoted by '$\lim$' in the following), a finite 
fraction of the total number of vertices belongs to this cluster. 
Therefore, we get the results : 
$m_1' = 1$ for $p < p_{cr}$ and $m_1' < 1$ for $p > p_{cr}$. Moreover, 
$m_1'$ is a decreasing function of the occupation probability. 
This typical behaviour is commonly (and incorrectly) 
called : '{\it the failure of mass conservation}', but, as stated before,
$m_1'$ is more simply the probability for a vertex to belong to some finite 
cluster.

On the other hand, the infinite set of Smoluchowski equations \cite{smo} :
\begin{eqnarray}
\label{eq3}
\frac{dc_s}{dt} = \frac{1}{2} \sum_{i+j=s} K_{i,j} c_i c_j -
\sum_{j} K_{s,j} c_s c_j ~~~ \ ,
\end{eqnarray}
are the coupled non-linear differential equations in the variables 
$c_s$, i.e., in the concentrations of clusters of mass $s$. The time $t$ 
includes both diffusion and reaction times, and these equations
suppose irreversibility of aggregation, i.e.,  the cluster fragmentation is not
allowed. The coefficients $K_{i,j}$ represent the probability
of aggregation between two clusters of mass $i$ and $j$ per unit of time.
Some of them are explicitely known for different experimental
conditions \cite{simons}. But all such known aggregation kernels have the 
remarkable homogeneity feature : 
\begin{eqnarray}
K_{ai,aj} = a^\lambda K_{i,j}~~~ \ ,
\nonumber
\end{eqnarray}
for any positive $a$, with $\lambda$ called the homogeneity index. 
Maybe the simplest example of the
homogeneous kernel is $K_{i,j}=(ij)^\mu$. It has been 
shown theoretically that if $\mu$ is larger than 1/2, 
then there exists a finite
time, say $t_{cr}$, for which $m_1'$ becomes smaller than 1
for $t > t_{cr}$ \cite{gel}. This can be interpreted as the appearance of
an infinite cluster at the finite time $t_{cr}$, and it is 
tempting to put in parallel the occupation probability $p$
in the percolation model and the time $t$ in the Smoluchowski
approach. But we will see later that even if this parallel seems
reasonable ($p$ being the advancement of the aggregation process),
some physical quantities behave quite differently. Note at last
that in eq. (\ref{eq3}), as written above, the sum over
$j$ does not include the gel $j = \infty$ if any (since $c_{\infty } = 1/\infty
 = 0$~), so the reaction between sol and gel is not taken into account. 
In principle, an additional term realizing the sol-gel aggregation 
should be added\cite{ZEH} on the right-hand member of eq. (\ref{eq3}).

\section{Reversible and irreversible sol-gel transitions}

To see differences between both types of models, we shall
treat two particularly simple cases : the bond-percolation on the
Bethe-lattice, and the aggregation kernel $K_{i,j}=ij$ in the Smoluchowski
approach. They are
chosen to be as close as possible in the sense that clusters 
generated by the percolation model are branched structures (without any
loop) in an infinite-dimensional space. So, all their constituents
are at the surface and reactivity must then be proportional to their 
mass\cite{ZEH}.
If the diffusion of clusters is negligible, e.g., when all clusters
diffuse with the same velocity, or for highly concentrated systems, the
corresponding reactivity kernels $K_{i,j}$ between two such clusters
of mass $i$ and $j$ should be proportional to $ij$.

The bond-percolation on the Bethe-lattice with
coordination number $z$, has been 
solved by Fisher and Essam\cite{Fisher}. Here, the main result
we are interested in, is the average concentration\cite{Fisher} :
\begin{eqnarray}
\label{eq4}
c_s =z\frac{((z-1)s)!}{((z-2)s+2)! s!} p^{s-1} (1-p)^{(z-2)s+z} ~~~\ ,
\end{eqnarray}
and the first normalized moment :
\begin{eqnarray}
\label{eq5}
m_1'=(\frac {1-p}{1-p^{*}})^{2z-2}~~~\ ,
\end{eqnarray}                          
with $p^{*}$ being the smallest solution of equation :
\begin{eqnarray}
\label{eq6}
p^{*}(1-p^{*})^{z-2} =p(1-p)^{z-2} ~~~ \ .
\end{eqnarray}
Let us define $p_{cr} \equiv 1/(z-1)$. For $p < p_{cr}$, 
the only solution of the above
equation is : $p^{*} = p$, but when $p$ is larger than $p_{cr}$, then
there is a smaller non-trivial solution which behaves as 
$p_{cr}-|p-p_{cr}|$ near $p_{cr}$. Above this threshold,
the moment $m_1'$ is smaller than 1 and behaves approximately
as $1-2(p-p_{cr})/(1-p_{cr})$.
The marginal case $z=2$ corresponds to the linear-chain case.

Coming back to the concentrations, we can see that for large values
of the size $s$, the following Stirling approximation holds :
\begin{eqnarray}
\label{eq7}
c_s \sim s^{-5/2} \exp (-\alpha s)     ~~~ \ ,
\end{eqnarray}
with $\alpha $ given by :
\begin{eqnarray}
\label{eq8}
\alpha = \ln \left( \frac {p}{p_{cr}} (\frac {1-p}{1-p_{cr}})^{z-2} 
\right) ~~~\ .
\end{eqnarray}
For this model, a power-law behaviour of the concentrations
is seen at the threshold $p_{cr}$, where more precisely 
$c_s \sim s^{- \tau}$ with $\tau =5/2$. Outside of this threshold, 
an exponential cut-off is always present \cite{stau} . 
This sort of critical behaviour at equilibrium is 
analogous to the thermal critical phenomena, and in particular,
there exist two independent
critical exponents, for example $\tau $ and $\sigma $ which is the exponent
of the mean cluster-size divergence (here $\tau = 5/2$ and
$\sigma =1 $), to describe completely the critical features.

The case of the Smoluchowski equations is quite different. Putting
$K_{i,j} = ij$ in these equations, Leyvraz and Tschudi\cite{gel} 
showed that there exists a critical value of the time, say $t_{cr}$ (here :
$t_{cr} = 1$), such that the solution is :
\begin{eqnarray}
\label{eq9}
c_s & = & \frac {s^{s-2}}{s!} t^{s-1} \exp (-st) ~~~ {\mbox for}  
~~~t < t_{cr} \nonumber \\ \nonumber \\
c_s & = & \frac {s^{s-2}}{s!} \exp (-s)/t ~~~  {\mbox for} 
~~~ t > t_{cr} ~~~ \ ,
\end{eqnarray}
for size-distribution and :
\begin{eqnarray}
\label{eq10}
m_1'& = & 1 ~~~  {\mbox for}  ~~~ t < t_{cr} \nonumber \\ \nonumber \\
m_1' & = & 1/t ~~~  {\mbox for}  ~~~ t > t_{cr} ~~~ \ ,
\end{eqnarray}
for the first normalized moment. 

As explained in the previous section, the behaviour of $m_1'$
characterizes the sol-gel transition, but some other features
are interesting to compare to the percolation model.
Firstly, one can see that the power-law behaviour is present
for $t > t_{cr}$ and {\it not only at the threshold}, since for
large $s$, we have :
\begin{eqnarray}
\label{eq11}
c_s & \sim & s^{-5/2} \exp (-\alpha s) ~~~  {\mbox for}  
~~~ t < t_{cr} \nonumber \\ \nonumber \\
c_s & \sim & s^{-5/2} /t ~~~  {\mbox for}  ~~~ t > t_{cr} ~~~ \ ,
\end{eqnarray}
with $\alpha = t - t_{cr} - \ln(t/t_{cr})$.
The whole distribution for the finite-size clusters evolves 
similarly and the appearance of a power-law behaviour is
not a sign of the transition but rather a characteristics of
the gelation phase. Secondly, it has been proved that for more
general homogeneous kernels : $K_{i,j}=(ij)^{\mu }$~, there
exists a relation between the exponent $\tau $~ of the power-law behaviour
and the exponent $\sigma $~ of the divergence of the mean size, more
precisely : $\tau = \sigma +2$. Here for $\mu =1$~ we have $\tau =5/2$ 
and $\sigma =1/2$. So, just one exponent is needed to describe the complete
critical behaviour. 
In this sense, the reversible and irreversible sol-gel transitions, 
though close, are not equivalent. 

\section{The origin of fluctuations in reversible and 
irreversible sol-gel models}

As noticed by Einstein\cite{Einstein} , 
fluctuations of a macroscopic variable $M$~ at the equilibrium 
resembles a Brownian motion in the space of this variable and, hence,
one expects the fluctuations to verify :
\begin{eqnarray}
\label{eq12}
\frac {<(M-<M>)^2>}{<M>} \sim {\text {Constant}} ~~~ \ .
\end{eqnarray}                                          
'Constant' means here : independent of the mass of the system. This approach
should be true for any short-ranged correlations, i.e., far from any 
critical behaviour. 
When close to the critical point, fluctuations are correlated throughout the 
whole system and Ornstein - Zernike argument yields\cite{OZ} :
\begin{eqnarray}
\label{eq13} 
<(M-<M>)^2>~ \sim ~{\epsilon }^{\nu (d+ \eta -2)}    ~~~ \ ,
\end{eqnarray}
where $\epsilon$ is the
distance to the critical point, $d$ is the dimensionality of the
system, and $\nu $~, $\eta $~ are the two critical exponents related
respectively to the divergence of the correlation length and to the divergence 
of the correlation function. In the same way, 
the average value of $M$ behaves as 
$<M> \sim {\epsilon }^{\beta }$, so that, without making any additional  
assumption about relations between critical exponents, one obtains : 
\begin{eqnarray}
\label{eq14}
\frac {<(M-<M>)^2>}{<M>^ \delta } \sim {\text {Constant}} ~~~ \ ,
\end{eqnarray}                                                   
if and only if $\delta = \nu (d+ \eta -2)/ \beta $~. $\delta $ is here a free
parameter and should not be confused with yet another critical exponent. 
'Constant' in (\ref{eq14}) means : independent of $\epsilon $ for an infinite
system. On the other hand, but by trivial finite-size scaling, 
this must also be a 
constant independent of the mass $N$ of the system at the critical point. 
Moreover, if the standard relations between critical exponents hold, 
we obtain : $\delta = 2$ for any dimension $d$~ 
(for the Landau-Ginzburg theory 
we have to replace $d$ by the upper critical dimension $d_c=4$~, above
which the mean-field is valid). 
So the fluctuations of an order parameter in the thermodynamic systems 
are expected to behave differently at the critical point and outside of it.

The case of irreversible aggregation models is quite different. 
Fluctuations in off-equilibrium
physical processes are hard to analyze because they develop dynamically and,
moreover, they often keep memory of a history of the process. In this sense,
large fluctuations in such processes cannot be an unambiguous signature of the
critical behaviour. In theoretical studies, it is often assumed that the 
fluctuations are irrelevant for the correct description of the mean properties
of the system (e.g. in the Smoluchowski approach). 
This point will be revisited below.

We will focus on the cluster-mass distribution $n_s$~
, the cluster-multiplicity distribution $P(M_0')$, and 
the distribution of the gel-mass for finite systems. 
The cluster-multiplicity 
$M_0'$ is just the total number of clusters minus 1, i.e.,
the largest cluster is omitted. We wish to answer the following questions :
(i) what is the importance of the power-law behaviour of the mass-distribution 
in detecting the critical
behaviour?  and, (ii)  what are the scaling properties of multiplicity 
distribution and gel-mass distribution
near the sol-gel transition for reversible and irreversible models ?
The importance of the pertinent quantity $M_0'$ in this context 
is that it is directly 
observable in many experimental situations where the system is not too large and
the cluster masses are not directly accessible like , e.g., 
in the hadronization
process in strong interaction physics or in the process of  
formation of the large scale structures in the Universe.
Moreover, informations about the normalized moments $m_0'$ or their fluctuations
can also be obtained for large systems, such as in the polymerization, 
the colloid aggregation, or the aerosol coalsecence.

\section{Scaling of the multiplicity-distributions at the 
reversible and irreversible sol-gel transition}

The multiplicity distribution is intensely studied in the 
strong interaction physics
where simple behaviour of much of the data on hadron-multiplicity distribution
seems to point to some universality independent of the particular dynamical
process. Some time ago, Koba, Nielsen and Olesen\cite{koba} 
suggested an asymptotic
scaling of this multiplicity probability distribution in strong interaction
physics :
\begin{eqnarray}
\label{eq15}
<M_0'> P(M_0') = \Phi (z)~~ ,~~~~~ z \equiv \frac{M_0'-<M_0'>}{<M_0'>} ~~~ \ ,
\end{eqnarray}                                                         
where the asymptotic behaviour is defined as $<M_0'> \rightarrow \infty$, 
$M_0' \rightarrow \infty$ for a fixed \\ $M_0'/<M_0'>$ ratio. $<M_0'>$~
is the multiplicity averaged over an ensemble of independent events.
The KNO scaling means that data for different energies (hence differing
$<M_0'>$) should fall on the same curve when $<M_0'> P(M_0')$  is plotted
against the scaled variable $M_0'/<M_0'>$. Extending this assumption, 
we suppose the more general scaling form :               
\begin{eqnarray}
\label{eq16}
<M_0'>^{\delta } P(M_0') = 
\Phi (z_{\delta })~~, ~~~~~ 
z_{\delta } \equiv \frac {M_0'-<M_0'>}{<M_0'>^{\delta }} ~~~ \ ,
\end{eqnarray}
with $\delta $ a real parameter and $\Phi$ a positive function. This form 
will be called the $\delta $-scaling. KNO-case corresponds to $\delta = 1$.
The normalization of the probability distribution $P(M_0')$ and
definition of the average value of $M_0'$, provides the two constraints :
\begin{eqnarray}
\label{eq17}
\lim \int_{-<M_0'>^{1- \delta }}^{\infty} \Phi (u) du =1~~~ \ , \nonumber \\
\nonumber \\
\lim \int_{-<M_0'>^{1- \delta }}^{\infty} u \Phi (u) du =0~~~ \ ,
\end{eqnarray}                                                
which imply : $\delta \le 1$ since $\Phi $ is positive.

As shown by Botet et al \cite{latora}, 
the multiplicity distribution 
for the 3d-bond percolation model on the cubic lattice at the 
infinite-network percolation threshold exhibits the $\delta$ - scaling 
with $\delta = 1/2$~. Even though the system experiences the
second-order critical phenomenon, fluctuations of the multiplicity-distribution
remain small and the KNO-scaling does not hold. Of course, $m_0'$ is {\it not}
in this case an order parameter since $\tau > 2$ even though there is some
irregularity in its behavior passing the threshold. This non-analyticity
can be illustrated by the exact result for bond-percolation on the 
Bethe lattice. In this mean-field case, the normalized $0^{th}$-moment
is :
\begin{eqnarray}
\label{eq18}
m_0' = (1-\frac {z}{2} p^{*}) \left( \frac {1-p}{1-p^{*}} \right)^{2z-2} 
\simeq
\frac {z-2}{2(z-1)} -(z-1) \epsilon +(1-\frac {z}{2}) |\epsilon | ~~~ \ ,
\end{eqnarray}
with : $\epsilon = p-p_{cr}$, and $\epsilon \ll 1$. It is easy to see that 
there is a jump
of the first $p$-derivative of $m_0'$ : $-z/2$ for $p \rightarrow p_{cr}^{-}$,
and $(4-3z)/2$ for $p \rightarrow p_{cr}^{+}$. The proper
order parameter for this model is the normalized mass of the gel-phase, i.e.,
the mass of the largest cluster divided by the total mass of the system
$S_{max}/N$~. Different probability distributions $P(S_{max}/N)$~ 
for different system sizes $N$~ can be all compressed into a 
unique characteristic function (see Fig. 1) :
\begin{eqnarray}
\label{eq18a}
<S_{max}/N> P(S_{max}/N) = \Omega \left( \frac{S_{max} - <S_{max}>}{<S_{max}>}
\right)~~~\ ,
\end{eqnarray}
which is an analogue of the
the KNO-scaling function (\ref{eq15}) of the multiplicity probability 
distributions. This new result is important since it seems to be a
characteristic critical behaviour
of the order parameter. Now we will discuss what happens for a
dynamical transition. 
 
We have simulated Smoluchowski approach by a standard Monte-Carlo 
binary aggregation\cite{moi} . At each step of the event-cascade, a couple
of cluster, with masses $i$ and $j$, is chosen randomly with probability 
proportional to $(ij)^{\mu }$, the time is then increased by the inverse of
this probability, and the couple of clusters is replaced by a unique cluster
of mass $i+j$. The largest cluster is always taken into account in
this scenario. Let us discuss here the $(\mu = 1)$-case ($K_{i,j} = ij$).
With the proper normalization, the critical gelation time is 
$t_{cr} = 1/N$~,  where $N$~ is the total mass of the system. The power-law
size distribution is indeed recovered numerically with the right exponent
$\tau = 5/2$ (see Fig. 2). In contrast to the percolation model, 
the multiplicity
distribution shows $(\delta = 0.2)$-scaling (see Fig. 3) corresponding to
very small correlated fluctuations. The scaling function, which in this case 
is asymmetric and sharp, is also quite different
from the one for the percolation\cite{latora}~. 

From the point of view of the criticality, the moment $M_1'$ should be a better
candidate. The $M_1'$-distribution in the $z_{\delta}$ - variable with
$\delta = 0.67$~ for masses $N=1024$, $N=4096$ and $N=16384$ is shown
in Fig. 4. This
non-trivial value of $\delta $ is here a signature of the transition : 
it disappears for $t > t_{cr}$ (see Fig. 5). 
But this is not yet analogous to the KNO-scaling
because $M_1'$ is not exactly the order parameter. The true quantity
to use is the reduced average mass of the largest cluster : 
$S_{max}/N = (N-M_1')/N=1-m_1'$
for which $(\delta = 1)$-scaling holds (see Fig. 6), as suggested
by the phase-transition arguments, and in accordance also with the
equilibrium percolation model. 

The results we have obtained here for the reversible or
irreversible sol-gel transition in thermodynamical or dynamical systems 
are expected to satisfy the following general conjecture : 
\begin{itemize}
\item
The occurence of $(\delta = 1)$ - scaling in the probability 
distribution $P(M)$~
of a certain macroscopic quantity $M$ is the sign of a critical behavior,
with $M$ as the order parameter; 
\item
The occurence of scaling with $1/2 < \delta < 1$~ in $P(M)$~ is 
the sign of a critical behavior in the system with $M$ related to 
but not identical to the order parameter;
\item
The occurence of $\delta =1/2$ - scaling in the distribution $P(M)$~ 
is the sign that the variable $M$ is not singular.
\end{itemize}
These conjectures has also recently been checked for kinetic fragmentation 
systems\cite{botplo}, and confort them.

\section{Conclusions}

We have studied here the scaling features of certain macroscopic quantities
which are
relevant for the sol-gel transition, such as the two first moments of
the cluster-mass-distribution and the mass of the gel, i.e., the order
parameter. We have employed two different models :
the percolation model which exhibits the equilibrium phase transition 
and the off-equilibrium Smoluchowski theory which shows
a dynamical phase transition. The phase transition threshold in both cases
is associated with the very particular scaling (\ref{eq18a}) of the order parameter 
fluctuations. For Smoluchowski's equations, this scaling is seen uniquely
at the critical gelation time $t_{cr}$ and disappears both for $t < t_{cr}$ and
$t > t_{cr}$ . It is interesting to notice that for $t > t_{cr}$
, the cluster-mass-distribution
is a power-law but , nevertheless, the order parameter fluctuations exhibit the
small amplitude limit of fluctuations, i.e.,
 the Gaussian scaling. 

For quantities other than the order parameter, 
similar scaling laws (see eq. (16))  are present but now with the different 
value of the parameter $\delta $~ of the scaling law. We have shown that
the scaling : $<M>^{\delta } P(M) = \Phi ((M-<M>)/<M>^{\delta } )$~ for $M,<M>
\rightarrow \infty$, 
holds for any macroscopic quantity and the value of
$\delta $ is : $1/2$ for the non-critical quantities, $0 < \delta < 1$
($\delta \neq 1/2$ and $1$) for critical quantities which related but which are
non identical to the order parameter,
and $1$ for the order parameter, respectively. These new scaling laws could 
be used in the phenomenological applications either to find the phase transition
threshold or to investigate whether the observed quantity 
is : regular, 
critical or whether it corresponds to the order parameter in the studied
process. For example,
in high-energy lepton-lepton collisions, the scaling behaviour 
with $\delta =1$ has been reported in the multiplicity distributions of the
produced particles for various collision energies \cite{delphi} . This
bahaviour has been interpreted as a signature of the asyptotic limit 
(the Feynman
scaling) in the $S$-matrix bahaviour, in spite of the fact that the Feynaman
scaling has been experimentally disproved by the continuing rise of particle
density in the central region. The analysis presented in this work 
offers an alternative and much 
more plausible explanation for these findings in terms
of the off-equilibrium phase transition. 
It is indeed challenging to see the {\it a priori} 
unexpected relation between such different physical phenomena
as the sol-gel transtition phenomenon and the particle production in
the ultrarelativistic collisions of leptons. 
It would also be extremely interesting to apply the
mathematical tools proposed in this work 
in the more typical aggregation phenomena  as found
in ....

\vfill
 
\newpage
 
{\bf Figure captions}\\
 
{\bf Fig. 1} \\
The gel-mass distribution for the 3d-bond percolation 
model on cubic lattices of different sizes : $N=8^3$ (crosses), 
$N=10^3$ (asterisks), $N=12^3$ (triangles),
and for the fixed bond probability $p_{cr}=0.2488$. Each point 
is the average over $200000$ events.\\
 
{\bf Fig. 2} \\
Double-logarithmic plot of the average mass-distribution 
of the Monte-Carlo Smoluchowski
simulations with aggregation kernel $K_{i,j}=ij$, for the two different system
masses : $1024$ (diamonds) and 
$N=4096$ (filled circles)~,  at the critical gelation-time ($t_{cr} =1/N$). \\
 
{\bf Fig. 3} \\
The multiplicity distribution for the Monte-Carlo Smoluchowski
simulations with aggregation kernel $K_{i,j}=ij$, for the two different system
masses : $1024$ (diamonds) and $N=4096$ (filled circles), 
in the $z_{\delta}$-variable ( $\delta = 0.2$) ~,  at the 
critical gelation-time. Each point is the average over 
$10^5$ independent events.\\
 
{\bf Fig. 4} \\
The $M_1'$-distribution for the Monte-Carlo Smoluchowski
simulations with aggregation kernel $K_{i,j}=ij$, for the two different system
masses : $1024$ (diamonds) and $N=4096$ (filled circles), in the
$z_{\delta}$-variable ( $\delta =0.67$)~, at 
the critical gelation-time. Each point is average over 
$10^5$ independent events.\\
 
{\bf Fig. 5} \\
The same as Fig. 4, but in the $( \delta = 1/2)$-variable and
after the critical gelation-time ($t_{cr}=2/N$).\\
 
{\bf Fig. 6} \\
The $S_{max}$-distribution for the Monte-Carlo Smoluchowski
simulations with aggregation kernel $K_{i,j}=ij$, for the two different system
masses : $1024$ (diamonds) and $N=4096$ (filled circles), 
in the KNO-variable at the critical gelation-time. Each point is the 
average over $250000$ independent events.\\


\begin{references}
 
\bibitem{stauffer}
D. Stauffer, {\it Introduction to Percolation Theory}, (Taylor and Francis, 
London and Philadelphia, Penn., 1985); and references therein.

\bibitem{gennes} 
P.G. de Gennes, J. Phys. Lett. (Paris) {\bf L37}, 1 (1976); \\
D. Stauffer, J. Chem. Soc. Faraday Trans. {\bf II 72}, 1354 (1976).

\bibitem{smo}
M. von Smoluchowski, Z. Phys. Chem. {\bf 92}, 129 (1917); \\
R.L. Drake, in {\it Topics in Current Aerosol Research}, vol.3,
G.M. Hidy and J.R. Brock eds, (Pergamon Press, NY 1972).
 
\bibitem{fs}
W.H. Stockmayer, J. Chem. Phys. {\bf 11}, 45 (1943); and {\bf 12}, 125 (1944);\\
P.J. Flory, {\it Principles of Polymer Chemistry}, (Cornell University,
Ithaca, 1953).

\bibitem{simons}
S. Simons, J. Phys. A, {\bf 19}, L901 (1986).

\bibitem{gel}
F. Leyvraz and H.R. Tschudi, J. Phys. A {\bf 14}, 3389 (1981);\\
R.M. Ziff, E.M. Hendriks and M.H. Ernst, Phys. Rev. Lett. {\bf 49}, 593 (1981).

\bibitem{ZEH}
R.M.Ziff, M.H.Ernst and E.M.Hendriks, J. Phys. A {\bf 16}, 2293 (1983).

\bibitem{Fisher}
M.E. Fisher and J.W. Essam, J. Math. Phys. {\bf 2}, 609 (1961).

\bibitem{stau}
D. Stauffer, Phys. Rep. {\bf 54}, 1 (1979).

\bibitem{Einstein}  A. Einstein, Ann. Physik {\bf 19}, 371 (1906).

\bibitem{OZ}  L.S. Ornstein and S. Zernike, Proc. Acad. Sci. Amsterdam {\bf
17}, 793 (1914).

\bibitem{koba}
Z. Koba, H.B. Nielsen and P. Olesen, Nucl. Phys. {\bf B40}, 317 (1972);\\
A.M. Polyakov, Sov. Phys. JETP {\bf 32}, 296 (1971);
ibid. {\bf 33}, 850 (1971).

\bibitem{latora}
R. Botet, M. Ploszajczak and V. Latora, Phys. Rev. Lett. {\bf 78}, 4593 (1997).

\bibitem{moi}
R. Jullien and R. Botet, {\it Aggregation and Fractal Aggregates}, 
(World Scientific, Singapore 1987).

\bibitem{botplo}
R. Botet and M. Ploszajczak, to be published.


\bibitem{delphi}
P. Abreu et al. (DELPHI Collaboration), Phys. Lett. {\bf B 416}, 233 (1998);\\
M. Acciarri et al. (L3 Collaboration), Phys. Lett. {\bf B 371}, 137 (1996);\\
G. Alexander et al. (OPAL Collaboration), Z. Phys. {\bf C 72}, 191, (1996);\\
(L3 Collaboration), CERN-PPE/97-42; \\(OPAL Collaboration), CERN-PPE/97-15;\\
P. Abreu, Proc. XXVII Int. Symp. on Multiparticle Dynamics,
G. Capon, V.A. Khoze, G. Pancheri and A. Sansoni eds., Frascati, 1997, 
Nucl. Phys. {\bf B 71} (Proc. Suppl.), 164 (1999);\\
P. Abreu et al. (DELPHI Collaboration), Proc. ICHEP'98 Conf.,
Vancouver, 1998; DELPHI 98-16 CONF 117.


 
 
\end{references}
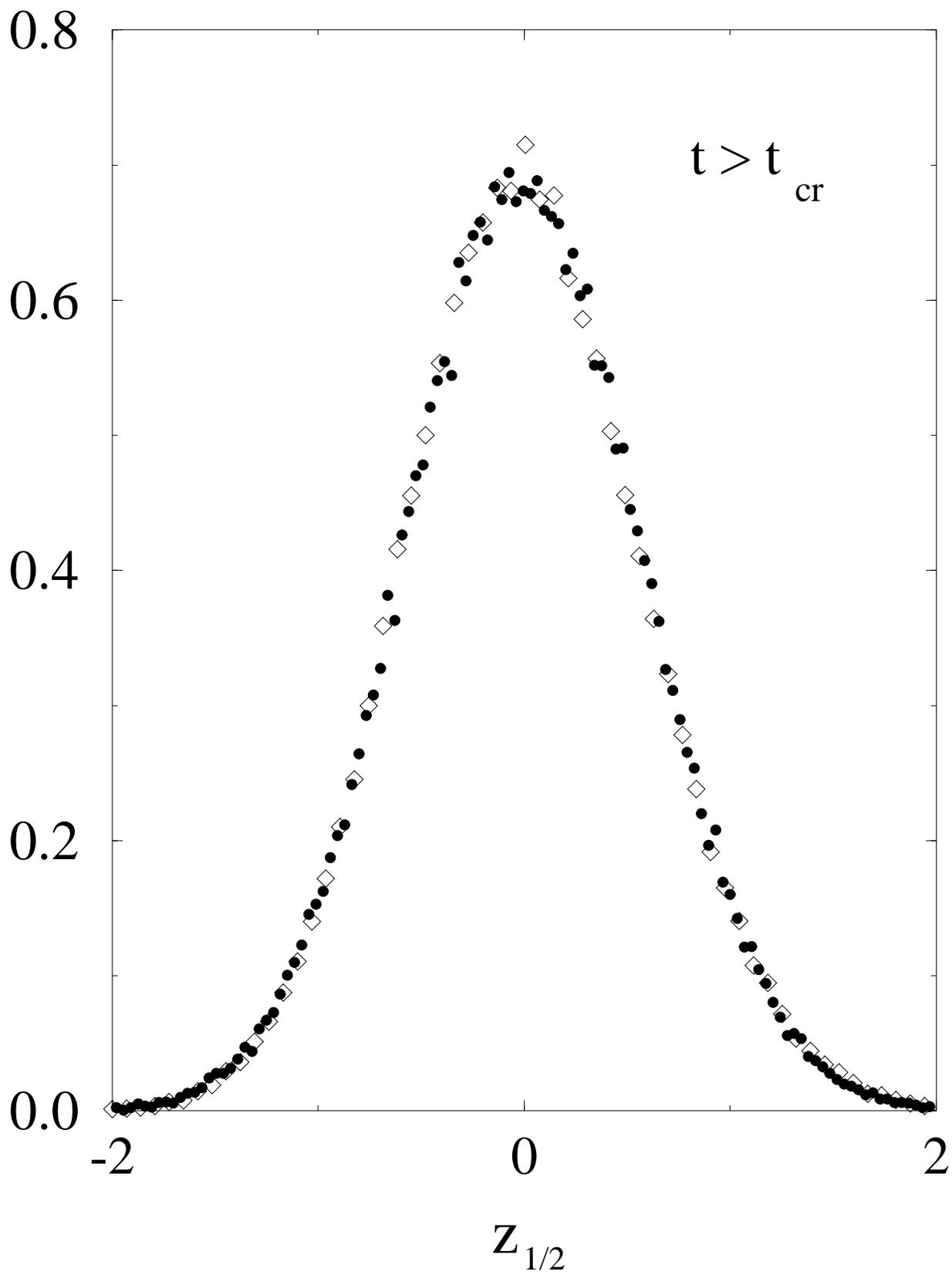
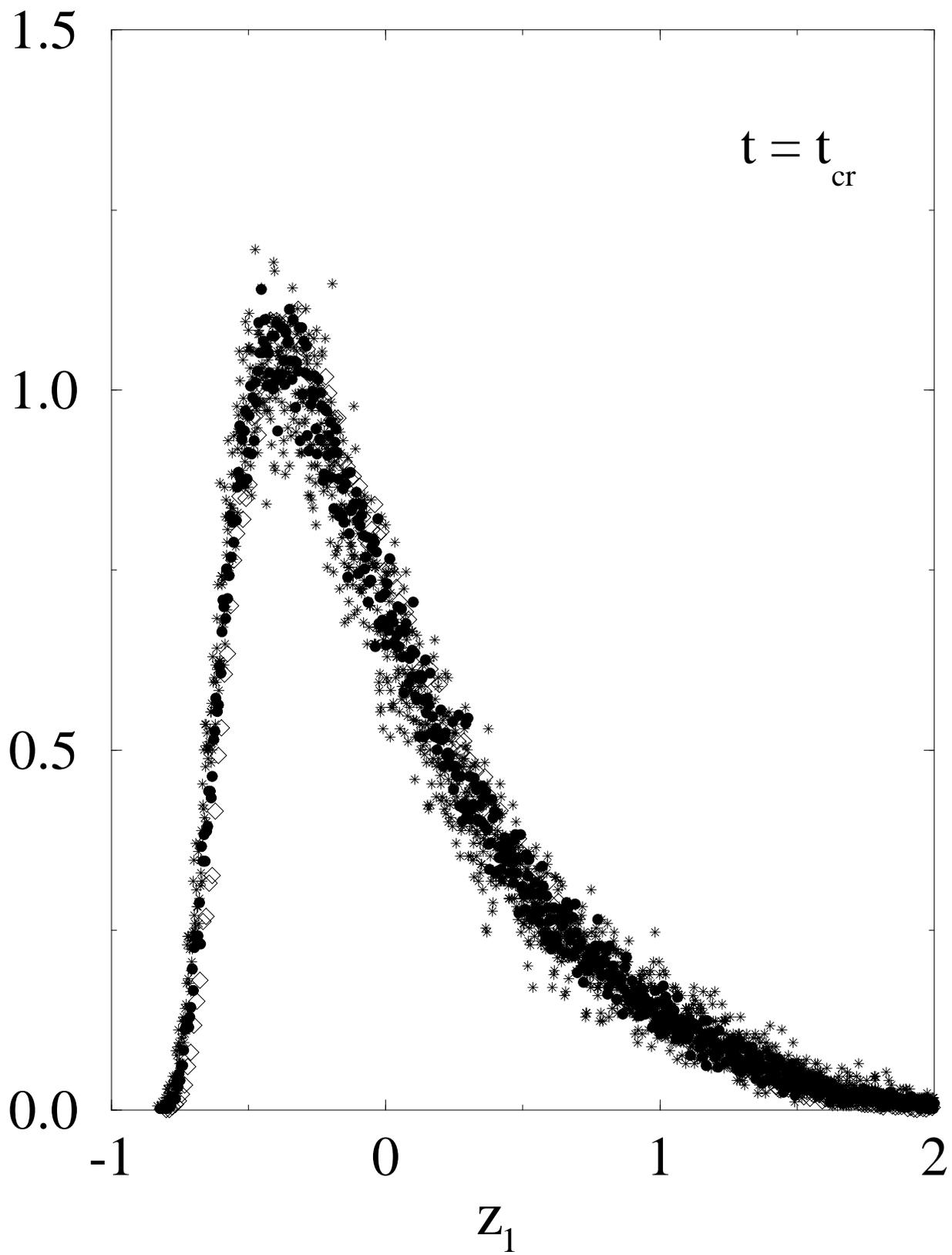
\end{document}